\documentstyle[prl,aps,epsf,floats]{revtex}

\newcommand{\ie}{{\em i.e.}\/ }
\newcommand{\half}{{1\over 2}}

\newcommand{\eg}{{\em e.g.}\/ }

\begin{document}
\draft

\twocolumn[\hsize\textwidth\columnwidth\hsize\csname@twocolumnfalse%
\endcsname

\title{
Anomalous scaling in depinning transitions}

\author{Onuttom Narayan}
\address{Department of Physics, University of California, Santa Cruz,
CA 95064}

\date{\today}

\maketitle

\begin{abstract}
It is demonstrated that the renormalization group (RG) flows of depinning 
transitions do not depend on whether the driving force or the system 
velocity is kept constant. This allows for a comparison between 
RG results and corresponding self-organized 
critical models. However, close to the critical point, scaling 
functions cross over to forms that can have singular behavior not seen 
in equilibrium thermal phase transitions. These can be different for the 
constant force and constant velocity driving modes, leading to different
apparent critical exponents. This is illustrated by comparing extremal dynamics 
for interface depinning with RG results, deriving
the change in apparent exponents. Thus care has to be exercised in such
comparisons.
\end{abstract}

]

Driven dynamics of disordered systems have been studied extensively
from two different approaches. As the external driving force is 
increased, a system undergoes a depinning transition from a 
macroscopically static state (with transient motion) to a moving steady 
state. This transition has been successfully described as a dynamical
second order phase transition~\cite{DSF}, and analyzed within the 
framework of the 
renormalization group (RG), originally for charge-density waves~\cite{CDW} 
and subsequently for interfaces~\cite{interf} and a variety of other 
systems~\cite{graball,ONAM}. This approach obtains the scaling behavior 
near the transition where the system starts to move. On the other hand, 
the concept of self-organized criticality (SOC), in the original sandpile 
model~\cite{BTW} and descendants thereof~\cite{PMB} has been used to obtain 
the behavior of systems that are forced to stay {\it at\/} the depinning transition. 

One would expect there to be connections between 
these two viewpoints~\cite{Tang}; indeed, it is possible to map automaton
models for CDWs to the original SOC model~\cite{ONAM,foot0}, and 
1+1 dimensional moving interface to a slightly different SOC model~\cite{Pac1}. 
The connections between the two have been exploited to obtain the 
dynamical critical exponent $z$ for CDWs by borrowing from exact 
results for SOC sandpiles~\cite{MD}. However, it has been pointed out~\cite{PMB} 
that one has to be careful whether the depinning transition is approached 
with a time independent external driving force, tuned to its critical
value, or with a time independent (infinitesimal) velocity. In the latter case, 
one envisages a feedback loop that adjusts the external force in a 
time-dependent way, so as to ensure that the rate at which the system 
moves (averaged over its entire spatial extent) is strictly time independent. 
It has been suggested~\cite{PMB} that the critical behavior at the 
depinning transition could be different for these two driving modes.
As an extreme example of constant current driving, there is the class of
``extremal models"~\cite{Sneppen,LesTang,TGR}, where at every (discrete) time 
step, there is activity only at one lattice site in the system.

For constant force driving, there are fluctuations in the (spatially averaged)
velocity of the system, while for constant velocity driving there are 
fluctuations in the external force. In either case, at a non-zero driving
rate, the fluctuations are small for sufficiently large systems, and thus
the two driving mechanisms should be equivalent. However, the large system limit 
is problematic at the critical
point. For instance, with constant force driving, the mean square 
fluctuations in the velocity of a large system are $(\delta v)^2\sim
v^2 (\xi/L)^d, $ where $v$ is the mean velocity, $\xi$ is the correlation
length at velocity $v$ (defined through the velocity autocorrelation 
function or other methods), $L$ is the linear extent of the system, and $d$
is its dimensionality. One would expect that holding the velocity constant
by adjusting the force will make a qualitative difference to the dynamics when
the velocity fluctuations are a significant fraction of the mean velocity,
\ie $\xi\sim L.$ Heuristically, for a fixed $v$ and $L\rightarrow\infty,$
imposing a constant velocity does not affect the dynamics, since the total 
`activity' at every time step is $v L^d,$ so that avalanches etc. 
are free to unfold with their own internal timescales. On the other hand, for fixed 
$L$ as $v\rightarrow 0,$ imposing a constant velocity `chokes off' the
dynamics: avalanches proceed sequentially, with internal dynamics determined 
by the velocity constraint. 
 
In this paper, we show that (to the extent that continuum descriptions are
appropriate) both constant force and constant velocity
driving actually have the {\it same\/} scaling under the renormalization
group. However, scaling functions for various quantities, which can 
have different behavior in the limits $\xi << L$ and $\xi >> L, $ 
are often different in the $\xi >> L$ regime for the two driving modes. 
It is standard for scaling functions in equilibrium critical phenomena 
to behave differently when their arguments are small and large, but 
as will be discussed later, dynamical critical points have even more 
freedom in how scaling functions can behave. It is this freedom that 
allows identical RG flows to still yield different scaling functions 
for the two driving modes in the $\xi >> L $ regime.

We demonstrate this by focusing on a particular system, that of a 
interface with short range internal elastic interactions
that is driven through a disordered medium under the 
influence of an external force. Due to the disorder, different parts
of the interface experience different random pinning forces as they
move forward.  Using the RG relationship between the 
interface velocity and the correlation length~\cite{interf} 
$v\sim\xi^{\zeta - z},$ we see that the crossover to $\xi> L$ occurs at
$v\sim L^{\zeta - z}.$ Note that extremal dynamics for a $d+1$ dimensional
interface corresponds to $v\sim L^{-d},$ and $z - \zeta < d$ in
all dimensions, so that $\xi >> L$ for large $L$.

It has been pointed out~\cite{SnepJen,LesTang,KTR}
that in extremal models for the motion of pinned interfaces,
the roughness of the interface can scale anomalously. Thus for an 
1+1 dimensional interface in a system of 
size $L$ in steady state~\cite{foot4}, if $h(x, t)$
is the interface position as a function of the transverse coordinate
$x$ and the time $t$, the roughness 
$w(t)=[\langle\{h(x, t + t_0) -  \overline h(t + t_0) - h(x, t_0) + 
\overline h(t_0)\}^2\rangle]^{1/2}$ 
(all averages are spatial averages) has the scaling 
form~\cite{KTR,foot1}
\begin{equation}
w(t) = t^\beta_v L^{- 1/2} \varphi\Bigg({L\over
{t^{1/z_v}}}\Bigg).
\label{scalform}
\end{equation}
(The subscripts on $\beta$ and $z$ denote constant velocity driving.)
It can be shown~\cite{LesTang,KTR} that the scaling function $\varphi$ goes to a 
constant for large values of its argument. The explicit $L$ dependence
that remains is anomalous, contrary to the normal expectation of a well defined
$L\rightarrow\infty$ limit. As a consequence of this, if the long 
time (steady state) roughness scales as $w(t\rightarrow\infty)\sim L^\zeta,$
then the conventional relation $\zeta = z \beta$ is replaced with 
\begin{equation}
\zeta = z_v\beta_v - \half.
\label{roughexp}
\end{equation}
With the additional result~\cite{LesTang,KTR,foot2}
\begin{equation}
z_v = 1 + \zeta
\label{dynexp}
\end{equation}
one can obtain $\beta_v$ in terms of $\zeta.$
Both Eq.(\ref{roughexp}) and Eq.(\ref{dynexp}) differ from 
renormalization group (RG) results~\cite{interf} for an interface 
driven with a constant applied force. In particular, the RG 
analysis yields {\it no\/} simple relation between the dynamic
exponent $z$ and the roughness exponent $\zeta,$ unlike
Eq.(\ref{dynexp}). Nor is there any reason to expect the 
anomalous scaling form of Eq.(\ref{scalform}), and therefore
$\zeta=z\beta$ should hold instead of Eq.(\ref{roughexp}).

We first demonstrate that, despite appearances, the RG flows are unaffected 
by going from a constant force driving mode to a constant velocity
driving mode. The standard equation for zero-temperature driven
dynamics of an interface is~\cite{KLT} 
\begin{equation}
\partial_t h(x, t) = \nabla^2 h(x, t) + Y(h(x, t); x) + F.
\label{FLR}
\end{equation}
Here $\nabla^2 h(x, t)$ comes from the (short-ranged) elastic energy of 
the interface, $Y(h; x)$ is a pinning force that comes from
a random (impurity) potential, and $F$ is the external driving
force. (Of course, it is not necessary that all lattice growth models
---\eg the Sneppen model~\cite{Sneppen}---can be mapped to this or any 
continuum equation.) 
We have neglected KPZ like terms~\cite{KPZ}. Using the 
Martin Siggia Rose method~\cite{MSR} and introducing 
an auxiliary field $\hat h(x, t),$ one can construct~\cite{interf} a 
generating functional $Z$ which can be written as 
\begin{eqnarray}
Z&=&\int [dh][d\hat h]\exp\Big[\int\!d^d x\, dt \nonumber\\
& &\qquad i\hat h(x, t)
\{\partial_t h - \nabla^2 h - F - Y(h; x)\}\Big]
\label{msr1}
\end{eqnarray}
where integrating out the auxiliary field yields a product of 
$\delta$-functions that imposes Eq.(\ref{FLR}). If instead the 
interface is driven at constant velocity, Eq.(\ref{msr1}) is 
replaced with 
\begin{eqnarray}
Z^\prime&=&\int [dF][d\mu][dh][d\hat h]\exp\Big[\int\!d^d x\, dt 
\nonumber\\
& &\qquad i\hat h(x, t)
\{\partial_t h - \nabla^2 h - F(t) - Y(h; x)\}\Big] \nonumber\\
& &\qquad\qquad+ i \mu(t)\{v - \partial_t h\}.
\label{msr2}
\end{eqnarray}
This extension is actually not difficult to understand: at any time
$t,$ by first integrating over $\mu(t)$ and $\hat h(x, t), $ we 
obtain $\delta$-function constraints that impose $\langle\partial_t h
\rangle = v$ (the average here is a spatial average) and 
Eq.(\ref{msr1}) with a (as yet unknown) driving force $F(t).$ 
Now performing the integral over $F,$ the integral together with the 
first constraint sets $F(t)$ to be whatever it has to be for 
$\langle \partial_t h$ to be equal to $v.$ One is left with the 
second constraint, \ie Eq.(\ref{FLR}), with $F(t)$ adjusted to 
ensure constant velocity. Even though $F(t)$ is now a dynamical
variable instead of a parameter, since it is not a field, \ie it
has no spatial dependence (nor has $\mu(t)$), when short distance
fluctuations are eliminated under renormalization there are no
extra loop corrections in $Z^\prime$ compared to $Z.$

As mentioned earlier, even though the RG fixed point is the 
same for constant force and constant velocity driving, the 
behavior of scaling functions are different for the two 
driving modes in the $\xi >> L$ regime.
To illustrate this, we first consider the scaling of the 
duration of an avalanche as a function of its linear size. 
This has the general scaling form 
\begin{equation}
t(l, L, \xi) = l^z T(l/L, \xi/L).
\label{avaltime}
\end{equation}
Here $l$ is the linear extent of the avalanche (in the $d$
transverse directions; in the direction the interface moves, the
extent is $\sim l^\zeta$), $L$ is the linear size of the system,
$\xi$ is the correlation length, and $z$ is the  non-trivial
dynamical exponent from the RG. For a fixed $l$ and $\xi,$ as 
$L\rightarrow\infty,$ the avalanche duration should be independent
of $L$ since the system is uncorrelated on 
length scales much bigger than $\xi$~\cite{foot3}.  It should also be independent 
of $\xi,$ since requiring a total growth rate of $v L^d$ for 
the system does not affect any individual avalanche from proceeding
with its own intrinsic timescale. Thus $T(0, 0)$ is some 
non-zero constant. 

In the other regime of $\xi>> L,$ \ie if the $v\rightarrow 0$ limit is 
taken {\it before\/} $L\rightarrow\infty,$ the avalanches are 
non-overlapping in time. In the constant velocity driving mode, where the 
velocity constraint is imposed, any single avalanche proceeds at a fixed rate. 
Therefore for the constant velocity driving mode, 
$t(l, L, \xi)$ must be inversely proportional to the rate at 
which the avalanche is allowed to proceed, which is $v L^d.$
Requiring that $t$ should have such a dependence on $v$ and $L,$
using the result~\cite{interf} $v\sim \xi^{\zeta - z},$ it is 
straightforward to verify that 
\begin{equation}
\lim_{\xi/L\rightarrow\infty} t(l, L, \xi) \sim 
l^{\zeta + d }/ (v L^d).
\label{avt}
\end{equation}
This yields an apparent dynamical exponent of $z_v = \zeta + d,$
which is the $d+1$ dimensional generalization of Eq.(\ref{dynexp}).
Note that there is {\it no\/} change in the apparent dynamical exponent in 
the $\xi >> L$ regime for constant force driving, where an avalanche
is allowed to proceed at its intrinsic rate (through parallel updating
of lattice sites). This is why the RG estimate for $z$ for two dimensional
CDWs agrees with the exact result for Abelian sandpiles~\cite{MD}.

The analysis of the interface roughness for a system of linear size L, 
$w(t, L),$ proceeds in a similar manner. At a velocity $v,$ the 
interface roughness has the scaling form 
\begin{equation}
w(t, L, \xi) = L^\zeta W(t/L^z; \xi/L).
\label{roughscal}
\end{equation}
For fixed $t$ and $\xi$, the roughness must have a well defined 
$L\rightarrow\infty$ limit: 
\begin{equation}
\lim_{L\rightarrow\infty} w(t, L, \xi) = \xi^\zeta W_1(t/\xi^z).
\end{equation}
For large $t,$ the roughness saturates to the steady state form
$w\sim \xi^\zeta.$ For $\xi\rightarrow\infty$ (the large $L$ limit 
has been taken first), or equivalently for small $t,$ the roughness
is $\xi$ independent, \ie $w\sim t^{\zeta/z}.$ 

In the other regime of $\xi>> L,$ the apparent exponents are once 
again different for constant velocity driving. In this case, 
the dynamics
are controlled by the imposed velocity, \ie the $v$ and $t$ dependence
of Eq.(\ref{roughscal}) occurs only in the combination 
$\tau = vL^d t.$ Using $v\sim\xi^{\zeta - z},$ this implies
\begin{equation}
\lim_{\xi/L\rightarrow\infty} w(t, L, \xi) = L^\zeta W_2(\tau/
L^{\zeta + d}).
\end{equation}
(With extremal dynamics, where one site moves forward at every time 
step, $\tau = t.$)
In the large time limit, the roughness approaches the steady state
form $w\sim L^\zeta, $ \ie $W_2(\infty)$ is a constant. It is not 
obvious how to extract the behavior of the function $W_2$ when its 
argument is small, but physical arguments supported by 
numerical results~\cite{LesTang,KTR} show that 
$w(\tau, L)$ must have a residual $L^{-d/2}$ dependence. Therefore
\begin{equation}
w\sim (v L^d t)^{\beta_v}/L^{-d/2}
\label{rghsc}
\end{equation}
with
\begin{equation}
\beta_v = {{\zeta + d/2}\over{\zeta + d}}
\end{equation}
which is the same as Eq.(10) of Ref.\cite{KTR} (generalized to 
$d$ dimensions).

Eqs.(\ref{avt}) and (\ref{rghsc}) give the $v\rightarrow 0$ 
scaling behavior of avalanche durations and interface roughness
respectively for constant velocity driving. We note once again that
these are for $\xi>> L,$ \ie $v\rightarrow 0$ before $L\rightarrow
\infty.$ As mentioned before, even in critical 
phenomena for equilibrium phase transitions, as one approaches
the transition for a fixed system size,  one sees a change in
the scaling form of dynamical variables. However, it is generally
possible to obtain the behavior in this regime by requiring that
there should be no dependence on (say) the reduced temperature $t$
for the behavior of a finite size system. This requirement comes
from the fact that there are {\it no} thermodynamic singularities for 
a finite sized system: it is possible to go smoothly from 
one side of the phase transition to another. No such requirement
exists for dynamical phase transitions, and one must
be careful about possible $v$-dependence even in the $v\rightarrow 0$
regime\cite{CDW}.

There are other mechanisms as well that can make the connection
between RG exponents and apparent scaling difficult. For instance,
for CDWs below the depinning threshold, the periodicity of the 
dynamical variable (the CDW phase) prevents it from advancing 
by more than $2\pi$ anywhere in a single avalanche, but the 
same periodicity makes a region that has just avalanched susceptible
to an imminent `retriggering' of a fresh avalanche~\cite{Middle}. 
The low 
frequency dynamics thus sees a non-trivial analog of the roughness
exponent, and the distribution of avalanche sizes is singular as
$v\rightarrow 0.$ Another example is for interface roughness 
itself, where the steady state roughness over a subsystem of 
size $x$ in a system with size $L$ scales as $ w^2(x; L)\sim 
x^2 L^{2\zeta - 2}$ when $\zeta > 1$~\cite{LL}. This is because 
the Fourier transform of the steady state roughness must be 
well behaved as $L\rightarrow \infty,$ and for $\zeta > 1$
$w^2(x; L)$ is dominated by long wavelength modes with $q\sim 1/L.$

In this paper we have shown that the renormalization group flows
for depinning transitions are the same whether the system is 
driven with a constant force or a constant velocity. However, in
the critical regime, the apparent scaling behavior of physical
quantities can be different for the two driving modes. This is 
because scaling functions can depend on the system velocity in 
a singular manner as the transition is approached. This is unlike
what is seen in equilibrium critical phenomena, and seems to be
more common in the constant velocity driving mode. Despite this
caveat, it is possible to obtain results for one driving mode
from the other.
 
I thank Anne Tanguy and Maya Paczuski for useful discussions.


\begin{references}

\bibitem{DSF}
D.S. Fisher, Phys. Rev. Lett. {\bf 50}, 1486 (1983); 
D.S. Fisher, Phys. Rev. B {\bf 31}, 1396 (1985).

\bibitem{CDW}
O. Narayan and D.S. Fisher, Phys. Rev. Lett. {\bf 68}, 3615
(1992); Phys. Rev. B {\bf 46}, 11520 (1992).

\bibitem{interf}
T. Nattermann, S. Stepanow, L-H. Tang and H. Leschhorn, J. Phys.
(France) II {\bf 2}, 1483 (1992); O. Narayan and D.S. Fisher,
Phys. Rev. B {\bf 48}, 7030 (1993).

\bibitem{graball}
D. Ertas and M. Kardar, Phys. Rev. E {\bf 49}, 2532 (1994);
J.P. Sethna and K. Dahmen, Phys. Rev. Lett. {\bf 71}, 3222 (1993) 
and Phys. Rev. B {\bf 53}, 14872 (1996); L. Balents, M.C. Marchetti
and L. Radzihovsky, Phys. Rev. B {\bf 57}, 7705 (1998);
P. Chauve, T. Giamarchi and P. le Doussal, Europhys. Lett. {\bf 44},
110 (1998) and cond-mat/0002299.

\bibitem{BTW}
P. Bak, C. Tang and K. Wiesenfeld, Phys. Rev. Lett. {\bf 59}, 381 (1987).

\bibitem{PMB} 
M. Paczuski, S. Maslov and P. Bak, Phys. Rev. E {\bf 53},
414 (1996) and references therein.

\bibitem{Tang} 
C. Tang and P. Bak, Phys. Rev. Lett. {\bf 60}, 2347 (1988);
C. Tang, K. Wiesenfeld, P. Bak, S.N. Coppersmith and P.B. Littlewood,
Phys. Rev. Lett. {\bf 58}, 1161 (1987); S.N. Coppersmith and 
P.B. Littlewood, Phys. Rev. B {\bf 36}, 311 (1987).

\bibitem{ONAM}
O. Narayan and A.A. Middleton, Phys. Rev. B {\bf 49}, 244 (1994).

\bibitem{foot0}
The mapping~\cite{ONAM} is from a CDW with periodic boundary
conditions to a sandpile~\cite{BTW} with periodic boundary conditions,
instead of the standard open boundary conditions. This 
creates an `above threshold' phase for the sandpile, where 
grains tumble around for ever without additional injection.  
However, avalanches below threshold are unaffected for a 
sufficiently large system. Another difference is that ramping up
the force below threshold for a CDW corresponds to dropping sand grains
on all the sites one by one in a random order, and cycling 
through this sequence repeatedly, instead of the standard spatially
{\it and\/} temporally random dropping of sand. Once again, this
does not matter in the large system limit.

\bibitem{Pac1}
M. Paczuski and S. Boettner, Phys. Rev. Lett. {\bf 77}, 111 (1996).
The mapping is from a SOC model to a 1+1 dimensional interface driven by
pulling it at one end at a constant speed. This is equivalent (after 
averaging over randomness) to driving the interface with a uniform 
applied force, with one end constrained to move at the (spatial and 
temporal) average velocity of the entire interface. The constraint 
is inconsequential for large systems.

\bibitem{MD} 
S. Majumdar and D. Dhar, Physica A {\bf 185}, 129 (1992).

\bibitem{Sneppen}
K. Sneppen, Phys. Rev. Lett. {\bf 69}, 3539 (1992). Only model B
of this paper will be considered.

\bibitem{LesTang}
H.~Leschhorn and L-H.~Tang, Phys. Rev. E {\bf 49}, 1238 (1994).

\bibitem{TGR}
A. Tanguy, M. Gounelle and S. Roux, Phys. Rev. E {\bf 58}, 1577 (1998).
This paper considers long ranged as well as (effectively) short ranged 
elastic interactions.

\bibitem{SnepJen}
K. Sneppen and M.H. Jensen, Phys. Rev. Lett. {\bf 71}, 101 (1993).

\bibitem{KTR}
S. Krishnamurthy, A. Tanguy and S. Roux, Eur. Phys. J. B {\bf 15},
149 (2000). 

\bibitem{foot4}
In Ref.\cite{KTR}, using the model of Ref.\cite{TGR}, the interface is in fact
started with a flat configuration, but the same results are obtained starting 
from steady state (A. Tanguy, private communication). For 
the model of Ref.\cite{Sneppen}, these results are {\it only\/} obtained starting
from steady state~\cite{SnepJen}. This is probably because the 
model of Ref.~\cite{Sneppen} is a `hybrid' model, where only a single site moves 
at a time step, but can then trigger a brief avalanche (at the same time step). 
This is neither constant velocity nor constant force driving. In steady state, 
the avalanches have a characteristic size, and the model effectively has constant 
velocity driving. However, when one starts from a flat interface, the 
mean avalanche size rises from unity to its saturated value; in fact, 
all scaling in the short time regime for this case is quite imperfect. (The 
connection between the changing mean avalanche size and the peculiar short time
behavior of the roughness is seen better in the original version of the 
model~\cite{Sneppen}, where the initial pinning strengths are Gaussian 
distributed, but the new pinning strengths when sites move are uniform over $[0, 1)$.)
Also, concepts of correlation lengths and compact avalanches are 
more tenuous~\cite{LesTang} in this model. 

\bibitem{foot1}
The scaling form looks slightly different than that in Ref.~\cite{KTR},
because the time $t$
is used and not the scaled time $\theta.$ The dynamic exponent $z_v$ is the
$z_2$ of Ref.~\cite{KTR}.

\bibitem{foot2} 
Numerical estimates $z_v$ and $\zeta$ for interfaces with short range 
elasticity (see Eq.(\ref{FLR})) in 1+1 dimension violate this equation, but 
this is probably because of pathologies for $\zeta > 1.$ Numerical estimates 
for long range elastic models in 1+1 dimension satisfy this equation~\cite{TGR}.

\bibitem{KLT}
R. Bruinsma and G. Aeppli, Phys. Rev. Lett. {\bf 52}, 1547 (1984); J. Koplik
and H. Levine, Phys. Rev. B {\bf 32}, 280 (1985); D.A. Kessler, H. Levine and
Y. Tu, Phys. Rev. A {\bf 43}, 4551 (1991).

\bibitem{KPZ} 
M. Kardar, G. Parisi and Y-C. Zhang, Phys. Rev. Lett. {\bf 56}, 889 (1986).
Such terms exist and are relevant for an interface moving in an anisotropic
medium, where the pinning has different strength in different 
directions~\cite{KTD}.

\bibitem{KTD}
L.A.N. Amaral, A.-L. Barabasi and H.E. Stanley, Phys. Rev. Lett. {\bf 73}, 
62 (1994);
L-H. Tang, M. Kardar and D. Dhar, Phys. Rev. Lett. {bf 74}, 920 (1995).

\bibitem{MSR} 
P.C. Martin, E. Siggia and H. Rose, Phys. Rev. A {\bf 8}, 423 (1973); 
H.K. Janssen, Z. Phys. B {\bf 23}, 377 (1976).

\bibitem{foot3} 
Avalanches can actually not be clearly separated in this regime, but it may 
be possible for continuum models to do this approximately by separating out 
fast and slow motion. In automaton models, this is much more 
problematic~\cite{Pac2}, but in any case, the discussion here should be valid for 
time scales corresponding to length scales for any measure of activity.

\bibitem{Pac2}
A. Corral and M. Paczuski, Phys. Rev. Lett. {\bf 83}, 572 (1999). For the 
model of Ref.~\cite{Pac1}, the avalanches are found to merge into each other
for strong driving. The scale found for the crossover from separate to merged 
avalanches is in fact just what one would expect by requiring $\xi\sim L$ and using 
numerically measured exponents for 1+1 dimensional depinning: 
$x = 1/(z - \zeta) = 0.205\pm 0.015,$ compared to~\cite{Lesch}
$\zeta = 1.25\pm 0.01$ and $\zeta/z = 0.88\pm 0.02.$ Other exponents 
in this paper also agree with results for 1+1 dimensional 
depinning~\cite{Lesch}.
 
\bibitem{Lesch} 
H. Leschhorn, Physica A {\bf 195}, 324 (1993).

\bibitem{Middle} 
A.A. Middleton, Ph.D. thesis, Princeton University, 1990; see also Ref.\cite{ONAM}.

\bibitem{LL} 
H. Leschhorn and L-H. Tang, Phys. Rev. Lett. {\bf 70}, 2973 (1993).

\end{references}
\end{document}